\documentclass[onecolumn,english,groupedaddress, superscriptaddress,nofootinbib,aps,prl,10pt,floatfix,notitlepage]{revtex4-2}

\usepackage[T1]{fontenc}
\usepackage[utf8]{inputenc}
\usepackage{babel}
\usepackage{amsmath,amsthm, amssymb}
\usepackage{graphicx}
\usepackage{url}
\usepackage{appendix}

\begin{document}

\title{Lockdown effects on air quality in megacities during the first and second waves of COVID-19 pandemic}

\author{Aswin Giri J}
\affiliation{Indian Institute of Technology Madras, India}

\author{Benjamin Sch\"afer}
\affiliation{Institute for Automation and Applied Informatics, Karlsruhe Institute of Technology, 76344 Eggenstein-Leopoldshafen, Germany}
\affiliation{School of Mathematical Sciences, Queen Mary University of London, United Kingdom}

\author{Rulan Verma}
\affiliation{Indian Institute of Technology Delhi, India}

\author{Hankun He}
\affiliation{School of Mathematical Sciences, Queen Mary University of London, United Kingdom}

\author{S. M. Shiva Nagendra}
\thanks{snagendra@iitm.ac.in}
\affiliation{Indian Institute of Technology Madras, India}

\author{Mukesh Khare}
\affiliation{Indian Institute of Technology Delhi, India}
\author{Christian Beck}
\affiliation{School of Mathematical Sciences, Queen Mary University of London, United Kingdom}
\affiliation{The Alan Turing Institute, 96 Euston Road, London NW1 2DB, UK}

\begin{abstract}
Air pollution is among the highest contributors to mortality worldwide, especially in urban areas. During spring 2020, many countries enacted social distancing measures in order to slow down the ongoing COVID-19 pandemic. A particularly drastic measure, the “lockdown”, urged people to stay at home and thereby prevent new COVID-19 infections during the first (2020) and second wave (2021) of the pandemic. In turn, it also reduced traffic and industrial activities. But how much did these lockdown measures improve air quality in large cities, and are there differences in how air quality was affected? Here, we analyse data from two megacities: London as an example for Europe and Delhi as an example for Asia. We consider data during first and second wave lockdowns and compare them to 2019 values. Overall, we find a reduction in almost all air pollutants with intriguing differences between the two cities except Delhi in 2021. In London,  despite smaller average concentrations, we still observe high-pollutant states and an increased tendency towards extreme events (a higher kurtosis of the probability density during lockdown) during 2020 and low pollution levels during 2021. For Delhi, we observe a much stronger decrease of  pollution concentrations, including high pollution states during 2020 and higher pollution levels in 2021. These results could help to design policies to improve long-term air quality in megacities.
\end{abstract}

\maketitle

\section{Introduction}

Numerous environmental challenges are faced by cities around the world, and air pollution is among the most pressing topics\citep{WHOAirPollution}. Citizens are forced to breathe air of persistent low quality and it  has a detrimental impact on their health and well-being \citep{molina2004megacities,kumar2015new}.
Overall, air pollution is linked to various diseases, like lower respiratory infections, strokes, cancers, asthma attacks, coughs, and chronic obstructive pulmonary diseases \citep{Amster, Shah, USEPA, guttikunda2013health, WHOPollution}. As per the State of Global Air 2019 \citep{hei2019state}, long-term exposure to outdoor and indoor air pollution contributed to nearly 5 million deaths in 2017. Out of these, 3 million deaths are directly attributed to particulate matter of 2.5 microns or smaller ($PM2.5$). Furthermore, high pollutant concentrations also damage the environment \citep{Dom, Astier, Must}.

\begin{figure}
    \centering
    \includegraphics[width=1\columnwidth]{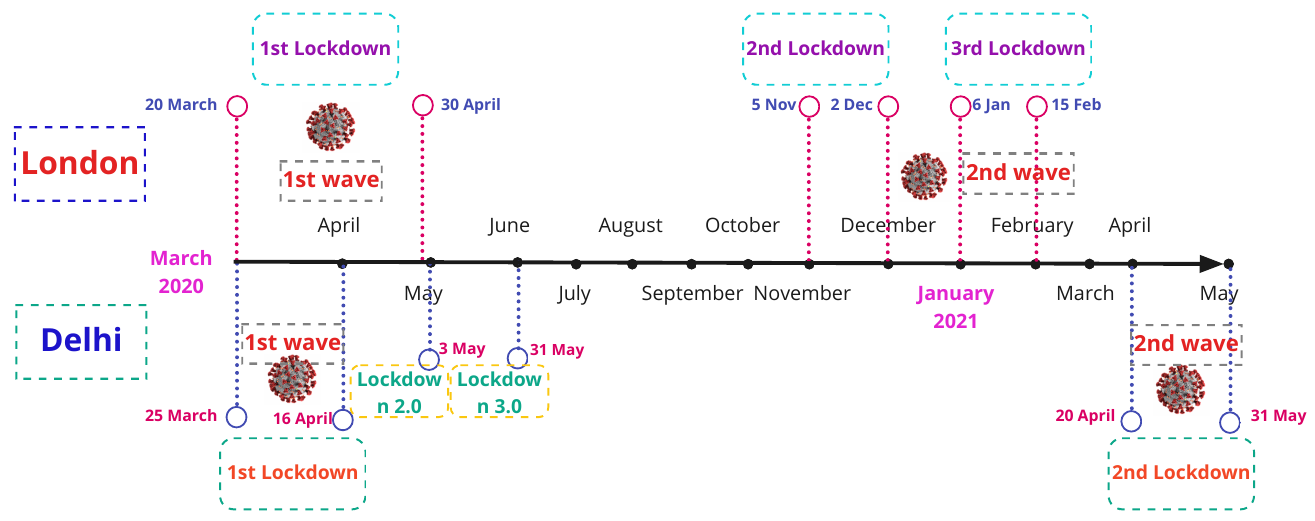}
    \caption{Lockdown timeline due to first and second waves of COVID-19 pandemic in London and Delhi}
    \label{fig:Timeline}
\end{figure}

During March 2020, the World Health Organisation declared the quickly spreading COVID-19 disease (caused by the SARS-COV-2 virus) as a global pandemic due to its rapid transmission and severe health effects. Many nations around the world were observing an alarming increase in the number of infected cases. To combat COVID-19, many countries initiated a “lockdown”, asking or forcing citizens to stay at home, leading to decreased mobility, increased social distancing and more working hours spent at home. During the COVID-19 lockdown, there was a significant reduction in air pollution levels across many countries. The situation is a unique opportunity to understand the baseline emissions in urban environment under lockdown conditions in different areas of the cities (suburban, traffic, and urban). Different governments imposed lockdowns of varying degrees at different time to reduce the spread of SARS-CoV-2 virus (Fig.~\ref{fig:Timeline}).

In this study, we focus on the analysis of air quality in two megacities, namely London as an example of an established Western megacity and Delhi as an Asian megacity in an emerging region. These two cities have very different properties when it comes to climate and pollution. London has a temperate oceanic climate, whereas Delhi features a dry-winter humid subtropical climate.

Delhi was identified as one of the world’s most polluted regions for PM2.5 in 2019 \citep{IQAir}, and during the winter months, the PM concentrations were observed to be 5 times higher than the annual averages due to stable meteorological conditions\citep{tiwari2018pollution}.
In contrast, air quality in London has improved in recent years as a result of policies to reduce emissions, primarily from road transport, such as Low and Ultra Low Emission Zones \citep{LondonLowEmission2020, LondonLowEmission2020, LondonPollutionHealth2020}. 
Further information on air quality in Delhi and London is provided in Supplementary Note 1.

Within this paper, we first give an overview of the air quality in both London and Delhi. Next, we compare measurements from the first lockdown period (March to April 2020) and second lockdown (Jan to Feb 2021 in London, and April to May 2021 in Delhi) with measurements from the previous year (2019). In particular, we analyse individual trajectories, probability distributions and also higher statistical moments. Then, we continue with a discussion of which sources cause which type of air pollution and how adequate guidelines could improve air quality in cities.
We conclude that air quality is much easily improved in emerging regions by taking regulatory actions, while Western cities can still profit from reduced traffic and should also investigate residential and background pollution.

In contrast to previous studies on air pollution during COVID-19 lockdown, such as \cite{LondonAir_analysis,sharma2020effect,mahato2020effect}, we investigate the detailed probability distributions of different pollutants, analysing 
various locations within the cities, while still noting a general decline in pollution levels. Also, we analyse higher statistical moments and compare two very different megacities in detail.
 
\section{Methods}

\subsection{Data overview}

To quantify the impact of the lockdown, we compare the “COVID-19-lockdown" in 2020 with the “business-as-usual" scenario from 2019. The reason to compare the
March-April 2020 values with the same period in 2019, instead of January to February 2020 is that there is a clear seasonal dependence in air quality, e.g. due to the efficiency of catalysts depending on ambient temperature \citep{williams_superstatistical_2020}. The first (2020) and second (2021) lockdown in Delhi occurred during similar summer months, whereas for London, the second lockdown (2021) happened during the winter month of January and February. We've used Ventilation Coefficient (V.C) to offer insights on the effect of seasonal change. 

\begin{figure*}
\begin{centering}
\includegraphics[width=0.9\columnwidth]{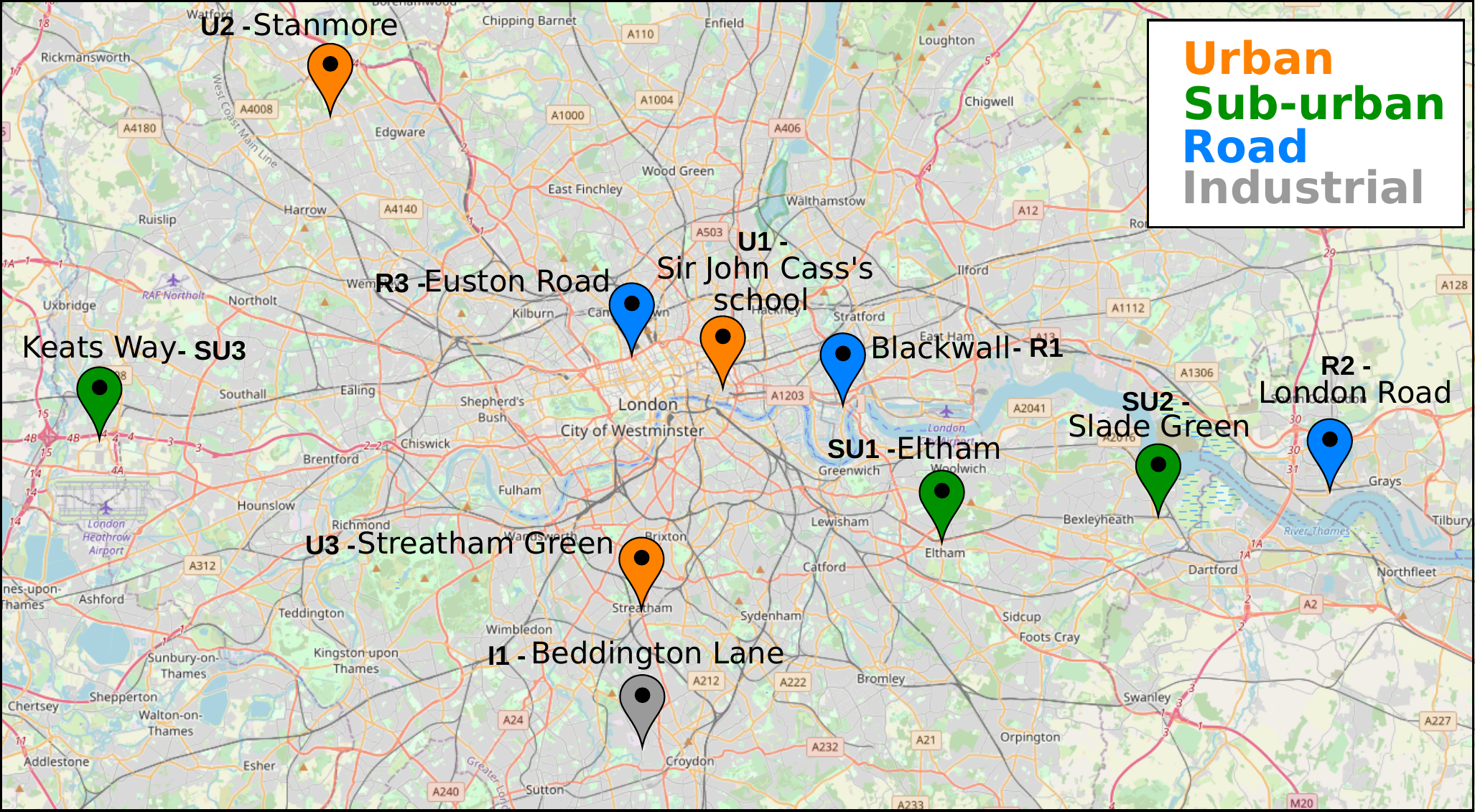}
\includegraphics[width=0.9\columnwidth]{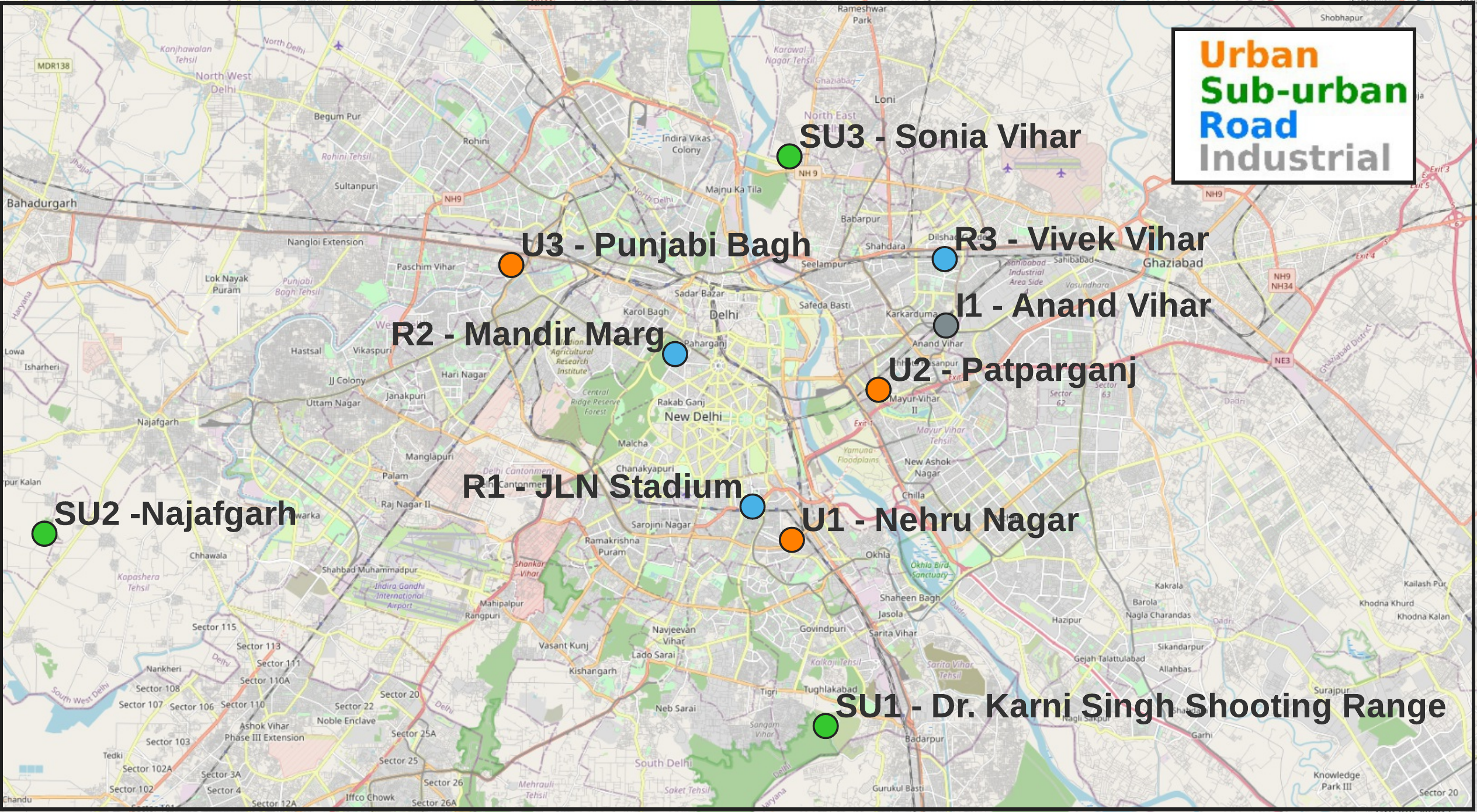}
\end{centering}
\caption{Measurement sites in London(a) and Delhi(b). We chose three urban, suburban and road locations each, as well as one industrial location. Map by openstreet maps.\label{fig:Map-Measurement-sites}}
\end{figure*}

For London, we utilize open data available from the London air quality network \citep{laqn}. From the available data, we select a total of ten locations for our analysis: three urban, suburban and road locations each and one industrial location (in the London data set very few industrial locations are available). The approximate locations are marked in Fig.~\ref{fig:Map-Measurement-sites}. As first lockdown period, we chose the dates from March 20 at 0:00 up to May 1 at 0:00. The UK closed off schools \citep{LockdownAnnouncement2020} on March 20 and went into a wider lockdown on 23 March. The second lockdown period was considered from 6 January 2021 at 0:00 up to 15 February 2021 at 0:00.

For Delhi, we utilize data provided by the Delhi Pollution Control Committee. From the available data, we again select three urban, suburban and road locations each and one industrial location. The exact locations of monitoring stations are marked in Fig.~\ref{fig:Map-Measurement-sites}. 
We analyse the main lockdown period between March 24 and April 21 and second main lockdown during the second wave was considered from April 20, 2021 to May 23, 2021.

We considered nitrogen oxides ($NO$ and $NO_{2}$  denoted as $NO_{x}$) as well as particulate matter, i.e. particulates of size less than 2.5 and 10 $\mu m$ ($PM2.5$ and $PM10$). Not only do $NO_{x}$ themselves have harmful impact on health, but they are also commonly used to indicate the presence of other pollutants \citep{Hamra}. In Supplementary Notes 2 and 3, we present further analysis in which $NO$ and $NO_{2}$ are analysed individually for the London data set, instead of being aggregated into $NO_{x}$. Unfortunately, ozone measurements, while found relevant in other cities \cite{sicard2020amplified}, were not available for all  measurements sites considered here, and have been omitted to maintain uniformity across the two cities. 

We use several road (R), urban (U) and suburban (SU) locations, with the following key:

For London we abbreviate: R1: Blackwall, R2: Thurrock, R3: Euston Road, U1: Sir John Cass School, U2: Stanmore, U3: Streatham Green, SU1: Eltham, SU2: Slade Green, SU3: Keats Way, I1: Beddington Lane.

For Delhi we abbreviate: R1: JLN Stadium, R2: Mandir Marg, R3: Vivek Vihar, U1: Nehru Nagar, U2: Patparganj, U3: Punjabi Bagh, SU1: Dr.Karni Singh shooting range, SU2: Najafgarh, SU3:Sonia Vihar, I1: Anand Vihar.

When plotting individual locations, we use the "1" index, i.e. R1, U1 and SU1 if not specified differently.

\subsection{Ventilation coefficient}
The ventilation coefficient (VC) indicates the ability of the atmosphere to disperse the pollutants and is computed as a function of the height of Planetary Boundary Layer (PBL) and wind speed\citep{prakash2017assimilative}, namely
\begin{equation}
    VC  = \mbox{PBL height} \; \times \; \mbox{Average Wind speed},
\end{equation}
where we typically measure height in meters, wind speed in meters per second and hence the VC in $m^2/s$.

To compare Delhi and London and 2019 with 2020, we obtained the approximate PBL height by using radiosonde data from the University of Wyoming \citep{AtmosphericDataWyoming2020} as follows. The PBL is the layer above the ground surface in which the pollutants are mixed and dispersed effectively. Right above the PBL is an inversion layer, which prevents the vertical movement of each air parcel. Here, we identify the PBL height as the lower boundary of this inversion layer, utilizing changes in potential temperature, relative humidity, moisture level, etc. The altitude corresponding to the maximum gradient of the potential temperature, mixing ratio and relative humidity profiles is taken as the PBL height similar to \cite{doi:10.1007/s10661-014-3615-y}.

\section{Time series}

\begin{figure}
\begin{centering}
\includegraphics[width=0.99\columnwidth]{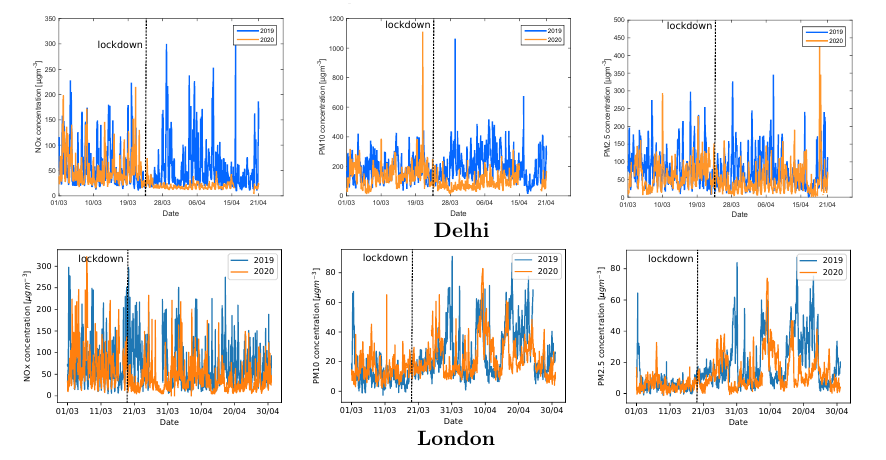}
\end{centering}
\caption{Pollution levels dropped substantially during the lockdown in mid-March 2020. We plot the trajectories of the concentrations of $NO_{x}$ (left), $PM10$ (center) and $PM2.5$ (right)  for Delhi (top row) and London (bottom row). We depict the urban location Punjabi Bagh for Delhi and the road location Blackwall for London. Dashed black lines indicate the approximate initiation of the lockdown in each city. \label{fig:Nox_PM10_trajectory}}
\end{figure}

We plot the temporal evolution of the pollution time series for individual locations and pollutants, to obtain an initial impression of the data. The pollutant concentrations display very large fluctuations but with a considerable drop in overall pollutant concentration after the 1st lockdown was initiated around  March 20, see Fig.~\ref{fig:Nox_PM10_trajectory}. The trend of decreasing pollutants is also observable for other pollutants (see Supplementary Notes 2 and 3). 
For Delhi, we observe a reduction in all pollutant levels after 25th March 2020. During the lockdown, we can see some instances of an increase in pollutant concentrations in early to mid-April. 
These may be due to the dust storms that occurred in Delhi during those days\citep{shandilya2007suspended,tiwari2012variations,tiwari2018pollution}. This effect can be clearly seen on the $PM10$ trajectory. Even though PM10 and NOx in Delhi showed similar trend after lockdown, PM2.5 showed large variability (Fig. ~\ref{fig:Nox_PM10_trajectory}). \cite{Dhaka2020} reported that low temperature and stagnation winds along with absence of solar radiation during early morning and evening hours resulted in formation of haze and mist. Even though there were no anthropogenic sources, there wasn’t any change in the diurnal variability of PM2.5. They suggested that surface level PM1 particles may grow with moisture and mix after the sun rise, causing large variability in PM2.5 concentration.

For London, we notice a  reduction of pollutants for the baseline $NO_{x}$ concentration (Fig.~\ref{fig:Nox_PM10_trajectory}). Furthermore, pronounced peaks are  visible in concentrations, both for $NO_{x}$ and for $PM10$. These peaks persist after the lockdown is enacted, and we will return to their systematic analysis via kurtosis values later. 

\section{Probability distributions}

\begin{figure}
\begin{centering}
\includegraphics[width=0.7\columnwidth]{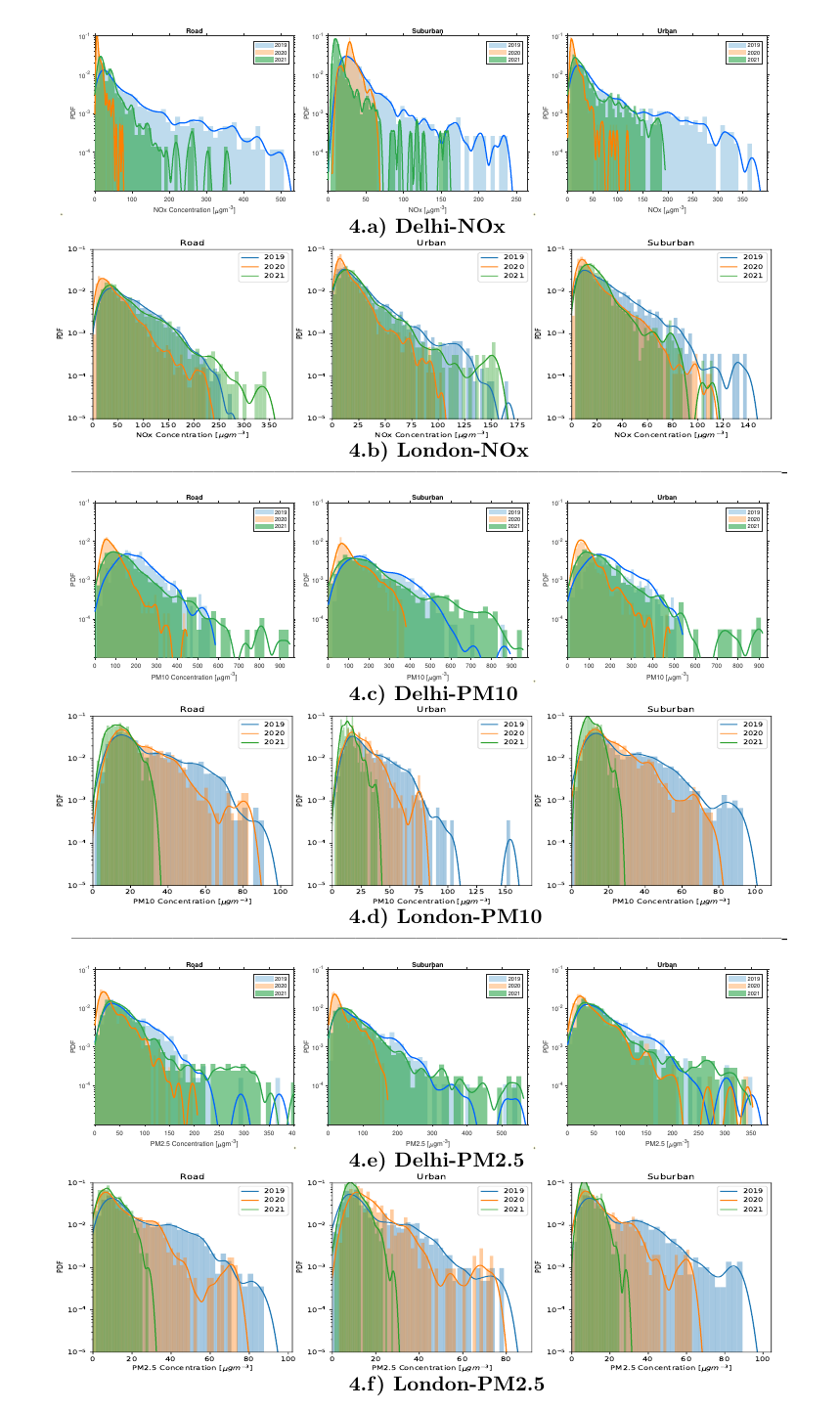}
\par\end{centering}
\caption{Probability density functions (PDFs) of the $NO_{x}$, $PM\,10$ and $PM\,2.5$  concentrations during 2019, 2020 and 2021. Top: $Delhi$. Bottom: $London$.  We plot both the normalized histogram and a Gaussian-Kernel estimate of the PDF. \label{fig:Distributions}}
\end{figure}

To investigate how likely certain pollution concentrations are reached, the empirical probability density functions (PDF) for $NO_{x}$, $PM 10$ and $PM2.5$ (Fig. \ref{fig:Distributions}) at an urban, a suburban and a road location were visualised. To this end, both the normalized histograms and a Gaussian-Kernel estimate of the empirical PDF were plotted \cite{waskom2020seaborn}.

The PDFs in 2019 tend to be broader than in 2020, i.e. reaching higher pollution states more frequently. Consistently, the 2020 distributions have a much more pronounced peak at low concentration levels. As expected, the pollution levels at the suburban location are generally lower than at the two other sites. The PDFs in 2021 are narrow and lower in London compared to 2019 and 2020, this may be due to the better meteorological conditions during Jan-Feb 2021. But in Delhi, the PDFs in 2021 are broader than 2019 for PM10 and PM2.5, even though they have similar meteorological conditions.

We note that the London distributions are all very similar in their peak near 0 concentration, with a following decay. In contrast, the Delhi data displays a maximum probability density at non-zero values, see e.g. the suburban measurement site. This might be explained by the different distributions observed when comparing $NO$ and $NO_{2}$ \citep{williams_superstatistical_2020}. In Supplementary Note 2, we further disentangle the impact of $NO$ and $NO_{2}$  for London.

\section{Moments}

To analyse the data more systematically, the first
and (normalized) forth moments of the empirical distributions were used. In particular, we compute the mean concentration $\mu=\frac{1}{N}\sum_{i=1}^{N}u_{i}$
and the kurtosis $\kappa=\frac{1}{N}\sum_{i=1}^{N}\left(\frac{u_{i}-\mu}{\sigma}\right)^{4}$, where $u_{i}$ is the pollutant concentration at step $i$, $\sigma$ is the standard deviation and $N$ is the number of measurements available. The kurtosis quantifies how many extreme events occur in the pollution concentration time series, i.e. how often high-pollution states are assumed. 
To exclude singular effects specific to one measurement site, we compute the moments for all ten measurement sites in both Delhi and London for 2019, 2020 and 2021.

The mean of $NO_{x}$  concentrations for all locations was lower in 2020 than it was in 2019. The observed drop in $NO_{x}$ concentrations were quite substantial within some locations, such as R3 in London, recording a decrease from more than 70 $\mu gm^{-3} $ down to merely 20 $\mu gm^{-3} $. This may be due to the reduced vehicular and industrial activities, leading to reduced emissions.
The same trend of decreasing mean values was also observed for particulate matter ($PM10$ in this case) in Delhi, but not for London. Notably, $PM10$ levels in Delhi reached four to ten times the values observed in London in the reference years. This is likely linked to background sources and non-human factors such as pollen or dust contributing substantially to $PM10$ concentrations. During lockdown, $PM10$ concentrations in Delhi drop substantially, almost reaching the same levels as in London.
During the 2021 lockdown in London, the concentration of PM10, PM2.5 and NOx have substantially decreased compared to 2019. This is due to the meteorological conditions being better suited for dispersion of pollutants. Conversely, in Delhi, during the 2nd lockdown the pollutant levels are almost equal to 2019, and in some cases exceeding the 2019 values.

In contrast to the mean, the kurtosis in 2020 and 2021 occasionally {\em exceeded} the values recorded in 2019 substantially. This observation is valid for $NO_{x}$, $PM10$ and $PM2.5$ and might be explained as follows: While on average the pollution levels were reduced, there were high-pollution states. These high pollution levels constitute extreme events under an otherwise reduced pollution level. Furthermore, their frequency occurrence contribute to a much higher kurtosis in 2020 than in 2019, where large pollution concentrations were more likely.
Interestingly, the kurtosis and hence the tendency to observe (local) extreme pollution states increased much more in London than in Delhi.

\begin{figure}
\begin{centering}

\includegraphics[width=0.93\columnwidth]{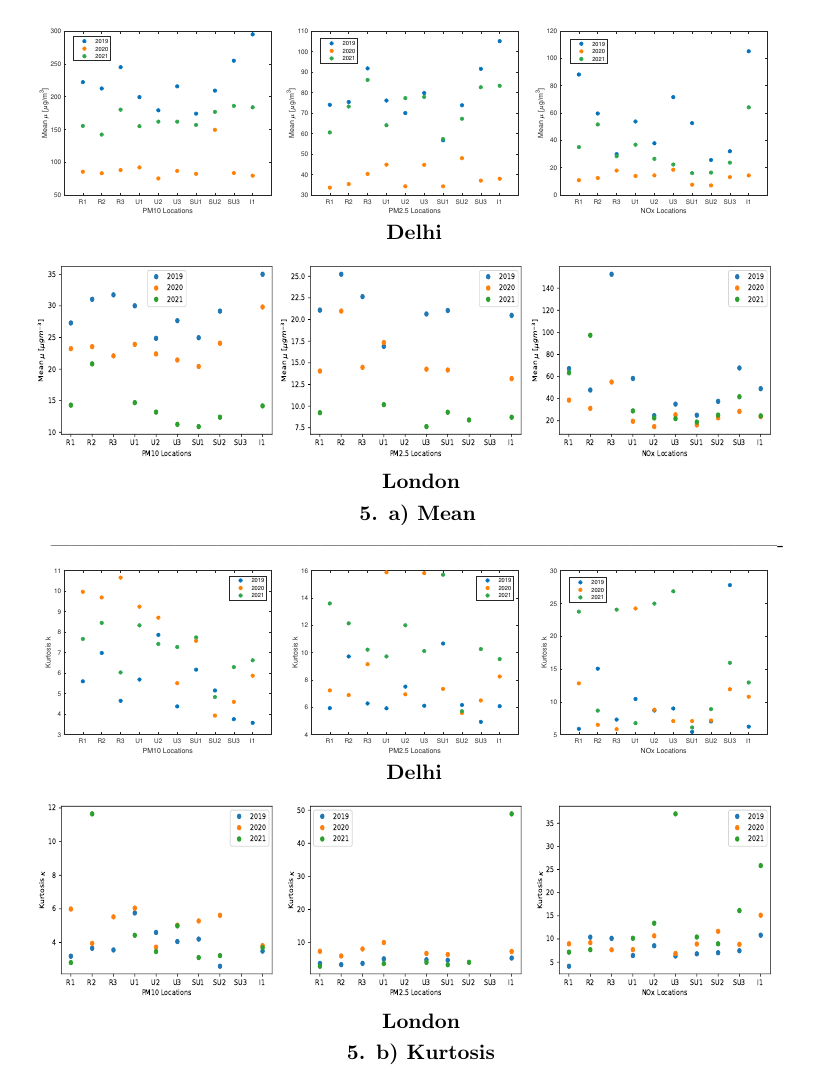}

\par\end{centering}
\caption{We compare the mean and kurtosis values from the lockdown period in 2020 and 2021 with 2019. Mean pollution levels dropped during lockdown, except Delhi in 2021 and the Likelihood of extreme events occasionally rose during lockdown. 
 Locations are abbreviated as road (R), urban(U), suburban (SU) or industrial(I). Top: $Delhi$ Bottom: $London$. Note the different y-axes scales. \label{fig:Compare-Mean-NO_PM_many_loc}}
\end{figure}

\section{Weather effects and high pollution states}

In this section, we answer two important questions: How much of the improved air quality could be attributed to weather effects, such as increased ventilation? Secondly, do high-pollution snapshots differ between 2021, 2020 and 2019?

\subsection{Meteorology and ventilation}
The Planetary Boundary Layer (PBL) was calculated from sounding data observed at 12 UTC. The mean PBL height was found to be 1940 m and 2200 m for London and Delhi, respectively. The typical PBL height varies from 1600 to 2200 m in London \citep{Bohnenstengel2015,Halios2018}and 2250 to 2700 m in Delhi \citep{2008atlas}.

Even though Delhi lies in a subtropical region and has a higher PBL height, its VC is lower than that for London due to relatively lower wind speeds (Table~\ref{tab:V.C}) \cite{DOI:10.1007/s12040-013-0270-6}. The National Meteorological Centre, USA and Atmospheric Environment Services, Canada, defined criteria for ventilation coefficients \citep{stackpole1967air,gross1970national}. The criteria for high pollution potential are $VC < 6000 m^2/s$ and mean wind speed $< 4 m/s$.  The dispersion potential is classified as low \citep{DOI:10.13140/RG.2.2.33949.82404} for $VC<2000 m^2/s$, medium  for $2000m^2/s<VC<6000 m^2/s$ and high for $VC>6000 m^2/s$. Thus, both the cities show a low dispersion and hence high potential for pollution during the 1st lockdown. Furthermore, the increased ventilation in London during 2020 has to be considered as small and cannot account for drastic changes in pollution levels. But during 2021 lockdown in London, the wind speed is high, enabling a better dispersion of pollutants and therefore reducing the pollutant concentration. Further analysis of wind statistics is given in Supplementary Note 4. 

\begin{table}[h]
    \centering
    \begin{tabular}{|c|p{1cm}|p{1cm}|p{1cm}|p{1cm}|p{1cm}|p{1cm}|}\hline
         \textbf{City} & \textbf{V.C 2019} & \textbf{V.C 2020} & \textbf{V.C 2021} & \textbf{W.S 2019} & \textbf{W.S 2020} & \textbf{W.S 2021}  \\ \hline
         \textbf{\emph{Delhi}} & 1451 & 1461 & 4537 & 0.62 & 0.68 & 2.31 \\ \hline
         \textbf{\emph{London}} & 1632 & 1808 & 9026 & 0.78 & 0.94	& 3.63 \\ \hline
    \end{tabular}
    \caption{
    Average Ventilation Coefficient (V.C), $m^{2}/s$  and Wind Speed (W.S), $m/s$ in Delhi and London}
    \label{tab:V.C}
\end{table}

\subsubsection{High-pollution snapshots}
\begin{figure}
\begin{centering}
\includegraphics[width=1\columnwidth]{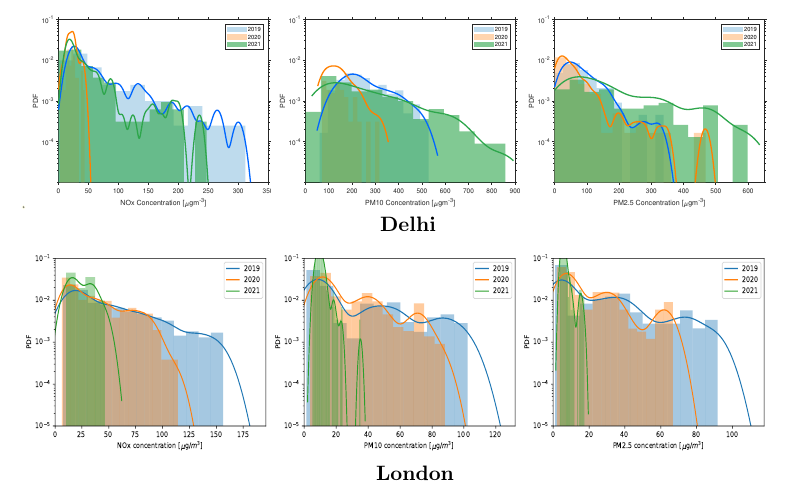}
\par\end{centering}
\caption{High-pollution states in Delhi became cleaner, but stayed almost constant in London in 2020. Much higher pollution levels were observed in Delhi and cleaner pollution level in London, during 2021 lockdown. We display high-concentration $NO_{x}$ snapshots, by comparing the 7-day period with the highest $NO_x$, $PM10$, \& $PM2.5$ variance between 2019, 2020 and 2021 in Delhi (Top) and London (bottom). Each data is evaluated at a suburban measurement site. Note again the log-scale of the y-axis. \label{fig:HighPollutionSnapshots}}
\end{figure}

Here, we compare typical high-pollution states in 2019, 2020 and 2021 to better understand the effect of the lockdown. We select data with a typical time window of length $T= 7$ days. Details of how such windows are selected based on super statistical approaches \citep{beck-cohen,BCS} can be found in \citep{williams_superstatistical_2020}. We select a snapshot within the 2019 data so that the variance of this snapshot is maximal, i.e. we select a local high-pollution state and then repeat this selection for 2020 and 2021. Analysing these high-pollution snapshots in Fig. \ref{fig:HighPollutionSnapshots}, we note that the data in 2020 can also reach high-pollution levels, almost as high as in 2019 (for London). However, these high pollution levels are less likely in 2020 than they were in 2019. PM2.5 high-pollution snapshots between 2019 and 2020 are almost identical. In 2021, there was very less chance of high pollution event in London, whereas in Delhi the high pollution levels were much more likely than 2019 (for PM10 and PM2.5).

\section{Attribution of pollutants emission}
This section discusses the attribution of the observed pollutant concentrations to individual emitters, and how they changed during the lockdown. 

Past studies estimated that about 75 per cent of PM pollution and $18\%$ of $NO_2$ in London originated from outside the city, while road transport is considered one of the main contributors of NOx ($50\%$) and PM10 ($53\%$) pollution. Contributing to road transport are the 3.98 million licensed motorized vehicles, which are mainly cars (3.2 million) \citep{RegisteredVehiclesLondon2020}. Therefore, to achieve clean air in London, both local and national policies are required \citep{SolvingAirQualityProblemLondon2015} to reduce emission from within and outside the city. 

In Delhi, air pollution arises from a variety of local and regional emission sources. The number of licensed vehicles in the city is 10.9 million, with a dominant share of two-wheelers (7.07 million) \cite{RegisteredVehiclesDelhi2018}. Past studies estimated that about $23 \%$ of the PM pollution and $36 \%$ of the $NO_x$ emission in the city were contributed from the road transport sector. About $52 \%$ of $NO_x$ emissions are attributed to industrial sources \citep{AirPollutionInDelhi2016}.  

For Delhi, the drastic reduction in pollutant concentrations could be attributed to a substantial reduction in vehicular and industrial activities. Not only did cars and industry emit fewer pollutants, but also less dust was re-suspended \citep{singh2020high}. Hence, for the future, road traffic regulations and dust resuspension should be monitored.
For London, traffic and industrial activities in the city and its surrounding decreased during the lockdown. Hence, air pollutant concentrations dropped, but the new mean distributions in $NO_x$ were higher than the lockdown $NO_x$ concentrations observed in Delhi. With road and industrial activities contributing less, it leaves the residential areas, the geographical surroundings and background sources \citep{LondonAir_analysis}. All of these should be monitored more thoroughly and regulations should be considered to improve air quality in the long term.

\section{Summary and Conclusions}

We compared the impact of COVID-19 induced lockdown measures on the air quality in two major cities. Mean pollution values tend to drop due to lockdown across all pollutants and for almost all investigated measurement sites. This holds for $NO_{x}$ in both London and Delhi, and for $PM$ in Delhi. 
While pollutant concentrations dropped both in London and Delhi, the reduction was much stronger in Delhi than in London. A specific observation for London is the change of the probability distributions, partially manifesting itself as an increase of kurtosis during lockdown. This is explained by the fact that temporary high-pollution states during and before the lockdown are not qualitatively different in London, but persist. In Contrast, not only did the mean drop, but the extremely polluted states are much rarer in Delhi during 1st lockdown.

Contrary to an earlier analysis for the London data \citep{LondonAir_analysis}, we did not observe a statistically significant increase in particulate matter (PM) concentrations during lockdown. In Delhi, there was a very substantial drop in PM concentrations. Note that we compared the spring 2020 season with the spring 2019 season, while \cite{LondonAir_analysis} compared it to the winter 2020. Hence, the increase in PM concentrations reported in \cite{LondonAir_analysis} might be a seasonal effect. 

Another study comparing the effect of the lockdown on different cities uses data based on daily air quality indices \citep{shrestha2020lockdown}. This is different from our study, where we make use of higher-resolved time series and also study higher moments systematically, such as the kurtosis.

The comparison between Delhi and London during the lockdown provides insightful lessons on air quality control:
A very strict lockdown in London did improve the air quality significantly, particularly in terms of NOx, highlighting the effectiveness of decreasing the traffic of vehicles. Simultaneously, it also shows that a very drastic improvement by regulating traffic or industry alone will not suffice, but pollution caused by other causes, such as residential or background, has to be taken into account as well.

The picture for Delhi is quite different: Without a lockdown, the pollutant concentrations are regularly 3 to 5 times as high as in London, indicating a much worse air quality in general. The 1st lockdown (2020) in Delhi improved air quality very drastically. This points to the great potential of clean air in Delhi if traffic and industrial emissions were reduced in the future by suitable control or regulation mechanisms.

There still remain many open questions, such as: Which sources in western and eastern cities can further be reduced to guarantee a long-term improvement in air quality? The lockdown
was at an enormous economic cost, but can a small change in behaviour
or a well-designed and balanced control mechanism lead to a sustainable and significant improvement in air quality in the future? 

Lockdown gave us many challenges and some information on baseline air quality levels. More insights about the basic atmospheric interactions sans anthropogenic intervention could be studied to widen our understanding of the natural atmospheric mechanisms.

\section*{Statements \& Declarations}

\subsection*{Acknowledgments}
The authors would like to thank the Delhi Pollution Control Committee (DPCC) New Delhi for their support and cooperation to this study.  
This project has received funding from the European Union’s Horizon 2020 research and innovation programme under the Marie Sklodowska-Curie grant agreement No 840825. We gratefully acknowledge funding from the Helmholtz Association under grant no. VH-NG-1727

\subsection*{Competing interests}
The authors declare no competing interests.

\subsection*{Author contributions}
B.S., S.N., M.K. and C.B. conceived and designed the research. B.S., A.G., R.V. and H.H. performed the data analysis and generated the figures. All authors contributed to discussing and interpreting the results and writing the manuscript. 

\subsection*{Data availability}
The code to reproduce the figures, along with the publicly available LondonAir data is also uploaded here: \url{https://osf.io/jfw7n/}

\bibliography{Reference}

\end{document}